\documentclass[a4paper,11pt]{article} 

\usepackage{jcappub}
\usepackage{bm}    
\usepackage{amsmath,amssymb}
\usepackage{color}  

\usepackage{comment}

\title{Improved calculation of the gravitational wave spectrum from kinks on infinite cosmic strings}
\author{Yuka Matsui,}
\author{Koichiro Horiguchi,}
\author{Daisuke Nitta}
\author{and Sachiko Kuroyanagi}

\affiliation{Department of physics and astrophysics, Nagoya University, Nagoya, 464-8602, Japan}

\emailAdd{matsui.yuka@f.mbox.nagoya-u.ac.jp}
\emailAdd{horiguchi.kouichirou@h.mbox.nagoya-u.ac.jp}
\emailAdd{nitta.daisuke@g.mbox.nagoya-u.ac.jp}
\emailAdd{kuroyanagi.sachiko@f.mbox.nagoya-u.ac.jp}

\abstract{Gravitational wave observations provide unique opportunities to search for cosmic strings.  One of the strongest sources of gravitational waves is discontinuities of cosmic strings, called kinks, which are generated at points of intersection.  Kinks on infinite strings are known to generate a gravitational wave background over a wide range of frequencies.  In this paper, we calculate the spectrum of the gravitational wave background by numerically solving the evolution equation for the distribution function of the kink sharpness.  We find that the number of kinks for small sharpness is larger than the analytical estimate used in a previous work, which makes a difference in the spectral shape. Our numerical approach enables us to make a more precise prediction on the spectral amplitude for future gravitational wave experiments. }  

\begin{document}
\maketitle

\section{Introduction}
Cosmic strings can naturally arise as a result of phase transitions
followed by spontaneously symmetry breaking in the very early universe
\cite{Kibble:1976sj,Vilenkin}.  It has also been argued that strings
of cosmological size can be formed in scenarios of the early universe
based on superstring theory, where they play the role of cosmic
strings \cite{Sarangi:2002yt,Jones:2003da,Dvali:2003zj}.  They are
among the few direct possible features of the very early universe and
could be employed to test high-energy theories.

Cosmic strings continuously generate gravitational waves throughout
the history of the universe after their formation.  Gravitational wave
bursts from different epochs and different directions overlap one
another and form a gravitational wave background over a wide range of
frequencies.  Thus, gravitational wave experiments are expected to be
a powerful tool to test the existence of cosmic strings.  Various types
of experiments can be used to probe the gravitational wave background
at different frequencies: pulsar timing experiments
\cite{Verbiest:2016vem,Janssen:2014dka} measure gravitational waves
at $\sim 10^{-8}$Hz; space missions such as eLISA
\cite{AmaroSeoane:2012km,AmaroSeoane:2012je} and DECIGO
\cite{Seto:2001qf,Kawamura:2011zz} explore $10^{-3}$Hz and $0.1$Hz,
respectively; ground-based experiments such as Advanced-LIGO
\cite{Harry:2010zz}, Advanced-VIRGO \cite{Accadia:2011zzc} and KAGRA
\cite{Somiya:2011np} focus on $\sim 100$Hz.

Gravitational wave signatures from cosmic strings have been
extensively investigated in the literature
\cite{Vilenkin:1981bx,Hogan:1984is,Sakellariadou:1990ne,Caldwell:1991jj,Caldwell:1996en}.
It has been widely accepted that the string network evolves towards
the scaling regime, where infinite strings continuously decay into
loops and the string network keeps ${\cal O}(1)$ infinite strings per
Hubble volume.  Thus, the network consists of infinite strings and
loops, both of which can be sources of gravitational waves.  In
refs. \cite{Damour:2000wa,Damour:2001bk}, it has been suggested that
non-smooth structures in strings, such as cusps and kinks, emit strong
gravitational wave bursts.  Cosmic string loops generically have cusps
and kinks, and various works have shown that they generate a large
gravitational wave background at high frequencies
\cite{Damour:2001bk,Damour:2004kw,Siemens:2006yp,DePies:2007bm,Olmez:2010bi,Binetruy:2010cc,Sanidas:2012ee,Sanidas:2012tf,Binetruy:2012ze,Kuroyanagi:2012wm,Kuroyanagi:2012jf,Sousa:2013aaa}.
While loops generate gravitational waves of wavelength shorter than
the loop size, gravitational waves from infinite strings become
important for long wavelength.  The spectrum of the gravitational wave
background originating from kinks on infinite strings are calculated
in ref. \cite{Kawasaki:2010yi}.

In this paper, we reexamine the spectrum of the gravitational wave
background from kinks on infinite strings.  Since the strength of
gravitational wave bursts depends on the sharpness of kinks, we need
to obtain the distribution function of the sharpness to calculate the
spectrum.  The evolution equation for the sharpness distribution is
modeled in ref. \cite{Copeland:2009dk}, and
ref. \cite{Kawasaki:2010yi} calculated the spectrum by using analytic
solutions of the differential equation for the distribution function.
The analytic solutions are obtained separately for radiation-dominated
(RD) and matter-dominated (MD) eras and the normalization for the RD
era is chosen to have the same amplitude with the MD era at
radiation-matter equality.  Instead of using analytic solutions, we
numerically solve the differential equation to obtain the sharpness
distribution function, which enables us to smoothly connect the RD and
MD eras.  In fact, since the string network evolves differently in these eras
\cite{Bennett:1987vf,Bennett:1989ak,Bennett:1989yp,Allen:1990tv,Vincent:1996rb,Vanchurin:2005pa,Ringeval:2005kr,Martins:2005es},
the parameters in the differential equation differ for MD and RD.
They should determine the normalization of the distribution function
and our numerical method correctly takes into account these effects.

The change of the parameters at radiation-matter equality is taken into account in two different ways.  First, we interpolate the values using the tangent hyperbolic function. Second, we calculate the time evolution of the parameters by using the velocity-dependent one-scale (VOS) model \cite{Martins:1996jp}. In the first case, the values of the numerical parameters are set to be the same as the previous work, which makes the comparison easier and enables us to show the effect of their change at radiation-matter equality clearly. The second case enables us to follow the scaling law of the string network and provides more realistic time evolution of the parameters.

The outline of this paper is as follows.  In section~\ref{sec:II}, we
briefly describe the methods to calculate the distribution function
for the kink sharpness and gravitational wave background spectrum.  In
section~\ref{sec:III}, we perform the numerical calculation to
evaluate the distribution function of kinks. Then, using the kink
distribution, we calculate the spectrum of the gravitational wave
background.  In section~\ref{sec:IV}, we make a comparison with
previous works.  Section~\ref{sec:V} is devoted to conclusions.

\section{Gravitational wave from kinks on the infinite strings}\label{sec:II} 
First, we review the dynamics of cosmic strings and describe the
method to calculate the distribution function of kink sharpness and
the power spectrum of the gravitational wave background.

\subsection{Dynamics of cosmic strings}
We consider cosmic strings in a spatially flat Friedmann-Lema{\^\i}tre-Robertson-Walker (FLRW)
metric,
\begin{equation}
  {\rm d}s^2 = a^2(\tau) \left (-{\rm d} \tau^2 +{\rm d} {\bm x}^2 \right ) = g_{\mu \nu} {\rm d} x^\mu {\rm d} x^\nu \, ,
\end{equation}
where $a(\tau)$ is the scale factor of the universe. A cosmic string is 
represented as
a two-dimensional worldsheet in the four-dimensional spacetime. We choose the coordinates on the worldsheet as $\zeta^1 = \tau$ (conformal time), $\zeta^2 = \sigma$ (a direction along a cosmic string), 
and $\frac{\partial x^\mu}{\partial \tau}\frac{\partial x_\mu}{\partial \sigma}=0$,
then the action of the Nambu-Goto string is given by
\begin{equation}
  S[x^\mu] = -\mu \int {\rm d}^2 \zeta \sqrt{-{\rm det} (\gamma_{ab})} \, , \label{eq:string_action}
\end{equation}
where $\mu$ is the tension of the string, 
$\gamma_{ab} = \frac{\partial x^\mu}{\partial \zeta^a}\frac{\partial x^\nu}{\partial \zeta^b}g_{\mu\nu}$
 is the induced metric on the string worldsheet.
 Taking the variation of the action with respect to $x^\mu$, we obtain the equation of motion for a cosmic string,
\begin{equation}
  \frac{\partial^2 {\bm x}}{\partial \tau^2} +\frac{2}{a} \frac{{\rm d} a}{{\rm d} \tau} \frac{\partial {\bm x}}{\partial \tau} \Biggl\{1 -\left (\frac{\partial {\bm x}}{\partial \tau} \right )^2 \Biggr\} = \frac{1}{\epsilon} \frac{\partial}{\partial \sigma} \left (\frac{1}{\epsilon} \frac{\partial {\bm x}}{\partial \sigma} \right ) \, , \label{eq:string_EoM}
\end{equation}
where
\begin{equation}
 \epsilon \equiv \sqrt{\frac{(\partial {\bm x}/\partial \sigma)^2}{1-(\partial {\bm x}/\partial \tau)^2}} \, ,
\end{equation} 
is interpreted as energy per unit $\sigma$, and we set $\epsilon=1$ at the present time.
When the Hubble friction is negligible, the equation has solutions of
left and right propagating waves. Accordingly, we define the new variable ${\bm
 p}_{\pm}$ which corresponds to the left and right moving modes,
\begin{equation}
  {\bm p}_{\pm} \equiv \frac{\partial {\bm x}}{\partial \tau} \mp \frac{1}{\epsilon} \frac{\partial {\bm x}}{\partial \sigma} \, . \label{eq:p_pm}
\end{equation}

\subsection{Cosmic string network}
Cosmic strings follow ``scaling law" where the number of infinite strings conserves in the horizon. 
In the VOS model \cite{Martins:1996jp}, the network evolution is characterized by the correlation length $L$. 
The total energy of a cosmic string and the average velocity are defined by
\begin{eqnarray}
  E & = & \mu a \int {\rm d} \sigma \epsilon \label{eq:total_energy} \\
  v^2 & \equiv & \frac{\int {\rm d} \sigma \left (\frac{\partial {\bm x}}{\partial \tau} \right )^2 \epsilon}{\int {\rm d} \sigma \epsilon},  \label{eq: string_v}
\end{eqnarray}
Then, the energy density $\rho_{\rm inf}$ of infinite strings is defined as 
\begin{equation}
  \rho_{\rm inf} = \frac{\mu}{L^2}.  \label{eq:infinite_string_E}
\end{equation}
Using the physical time $t$, which relates to the conformal time as ${\rm d} t = a {\rm d} \tau$, the evolution equations of the correlation length and velocity are 
\begin{eqnarray}
  \frac{{\rm d} L}{{\rm d} t} & = & HL(1+v^2) +\frac{1}{2}cpv,  \label{eq:Leq} \\
  \frac{{\rm d} v}{{\rm d} t} & = & (1-v^2) \left (\frac{k}{L} -2Hv  \right ),  \label{eq:veq}
\end{eqnarray}
where 
$k(v) \equiv \frac{1}{v (1 -v^2)}\frac{\int {\rm d} \sigma \bigl\{1-({\rm d} {\bm x}/{\rm d} \tau)^2 \bigr\} ({\rm d} {\bm x}/{\rm d} \tau) \cdot {\bm u} \epsilon}{\int {\rm d} \sigma \epsilon} \simeq \frac{2 \sqrt{2}}{\pi} \frac{1-8v^6}{1+8v^6}$ and ${\bm u}$ is a unit vector parallel to the curvature radius vector, and $H$ is the Hubble parameter $H = \frac{{\rm d} a/{\rm d} t}{a}$. The second term of the right hand of \eqref{eq:Leq} is the energy transmitted to loops per unit time, $p$ is a probability of reconnection and $c$ is the loop chopping efficiency parameter which is set $c \simeq 0.23$ \cite{Martins:2000cs}.
With $\gamma \equiv L/t$, the first equation is rewritten as 
\begin{equation}
  \frac{{\rm d} \gamma}{{\rm d} t} = \frac{1}{t} \Biggl\{-\gamma +H \gamma t(1+v^2) +\frac{1}{2}cpv \Biggr\} \, . \label{eq:gammaeq} \\
\end{equation}
By setting ${\rm d} \gamma/{\rm d} t$ and ${\rm d} v/{\rm d} t$ to be zero in
\eqref{eq:veq} and
\eqref{eq:gammaeq}, we obtain the asymptotic solutions
\begin{equation}
  \gamma = {\rm Const.}, \, v = {\rm Const.}\, 
\end{equation}
As we find the correlation length $L = \gamma t$ grows in proportion to $t$, the number of infinite strings is conserved in the horizon. The velocity keeps constant value for a fixed cosmic expansion rate.

\subsection{Distribution function of kinks on infinite strings} 
Kinks are defined as discontinuities in the string tangent vector ${\bm x}$. They are produced by reconnection between cosmic strings and propagate along strings.
The sharpness of the kink is defined by 
\begin{equation}
  \psi \equiv \frac{1}{2} (1- {\bm p}_{\pm, \, 1} \cdot {\bm p}_{\pm, \, 2}) \, . \label{eq:psi}
\end{equation}
The subscript $\pm$ denotes the left and right moving modes, and $1$/$2$ represent the left/right side of the discontinuity, respectively. The range of sharpness is $0 \leq \psi \leq 1$ and 
a large value of $\psi$ corresponds to a sharp kink.

Let us define $-\alpha \equiv \langle{\bm p}_+ \cdot {\bm p}_-\rangle=-(1-2v^2)$,
where the bracket means ensemble average in the string network, $v^2$ is the mean square velocity of strings.
Rewriting \eqref{eq:p_pm} and \eqref{eq:psi} in terms of $\psi$, we have  \cite{Copeland:2009dk} 
\begin{equation}
  \psi \propto t^{-2 \zeta} \, ,
\label{eq:blunting}
\end{equation}
where $t$ is the proper time $t=\int a d\tau$ and $\zeta = \alpha \nu$. The parameter $\nu$  characterizes the evolution of the scale factor as $a \propto t^\nu$. 
The value of $\zeta$ in the MD era differs from the one in the RD era, as we provide
 in table~\ref{tab:const}.

Intersections in the cosmic string network continuously generate kinks on infinite strings. We define 
the distribution function of kinks
 as a function of the sharpness and proper time, $N(\psi, \, t)$, so that $N(\psi, \, t)d\psi$ is the number of kinks between $\psi$ and $\psi +{\rm d} \psi$ within the volume $V$ at proper time $t$. Then its time evolution is given by \cite{Copeland:2009dk}
\begin{equation}
\frac{\partial N}{\partial t}(\psi, \, t) -\frac{2 \zeta}{t} \frac{\partial}{\partial \psi} (\psi N(\psi, \, t)) 
=\frac{\bar{\Delta} V}{\gamma^4 t^4} g(\psi) -\frac{\eta}{\gamma t} N(\psi, \, t),  \label{eq:dNdtwithoutY}
\end{equation}
where $\bar{\Delta}$ is the probability of the intersection\cite{Austin:1993rg}, 
$\gamma$ characterizes the correlation length of the string network $L$ as $L = \gamma t$, 
and $\eta$ is the decrease rate of kinks due to the loop production which is determined from simulations \cite{Kibble:1990ym}. 
The function $g(\psi)$ in \eqref{eq:dNdtwithoutY} is the initial sharpness distribution, and given by
\begin{equation}
  g(\psi) = \frac{35}{256} \sqrt{\psi} (15 -6 \psi -\psi^2) \, ,
\end{equation}
where we set $g(\psi) = 0$ for  $\psi<0$ or $1<\psi$.
When the left hand side of \eqref{eq:dNdtwithoutY} equals to zero,  the equation demonstrates that
the number of kinks is conserved while the sharpness decreases as in \eqref{eq:blunting}.
In the right hand side of \eqref{eq:dNdtwithoutY}, the first term denotes a production of kinks by intersection of strings, 
the second term denotes decreasing of the number of kinks by the loop production. This term can be obtained by considering the length of cosmic strings $d$ transferred from infinite strings to loops, 
\begin{equation}
\left. \frac{\dot{d}}{d} \right |_{\rm loop} = -\frac{\eta}{\gamma t} \, ,
\end{equation}
and we have assumed that the fraction of kinks taken away on loops is proportional to the loss of length, $\dot{d}/d \propto \dot{N}/{N}$. 

\begin{table}[htb]
  \begin{minipage}[t]{.45\textwidth}
    \centering
    \begin{tabular}{|c||c|c|} \hline 
      & RD & MD \\ \hline \hline
      $\gamma$ & 0.31 & 0.50 \\ \hline
      $\zeta$ & 0.09 & 0.2 \\ \hline
      $\bar{\Delta}$ & 0.20 & 0.21 \\ \hline
      $\eta$ & 0.18 & 0.1 \\ \hline
    \end{tabular}
    \caption{The values of the constant adopted in ref. \cite{Kawasaki:2010yi} are summarized for RD and MD eras.}
    \label{tab:const}
  \end{minipage}
  \hfill  
  \begin{minipage}[t]{.45\textwidth}
    \centering
    \begin{tabular}{|c||c|c|} \hline 
      & RD & MD \\ \hline \hline
      $\gamma$ & 0.27 & 0.56 \\ \hline
      $\zeta$ & 0.062 & 0.16 \\ \hline
      $\bar{\Delta}$ & 0.19 & 0.21 \\ \hline
      $\eta$ & 0.076 & 0.068 \\ \hline
    \end{tabular}
    \caption{The values of the constant for RD and MD eras obtained by solving the VOS equations.}
    \label{tab:const_scaling_solve}
  \end{minipage}
\end{table}

To obtain the kink distribution using (\ref{eq:dNdtwithoutY}), we need the time evolution of $\gamma, \, \zeta, \, \bar{\Delta},$ and $\eta$.  In this paper, we show results by using two different methods to obtain them.  In the first case, we use the parameter values used in ref. \cite{Kawasaki:2010yi} and we smoothly change them from RD to MD at radiation-matter equality $t_{\rm eq} \simeq 2.0 \times 10^{12} {\rm s}$ using 
\begin{equation}
  \chi(t) = \chi_m \frac{1+{\rm tanh}(100{\rm ln}(t/t_{\rm eq}))}{2} + \chi_r \frac{1-{\rm tanh}(100{\rm ln}(t/t_{\rm eq}))}{2} \, , \label{eq:tanh}
\end{equation}
where $\chi_m$ and $\chi_r$ describe values for MD and RD.  The values for RD and MD eras are listed in table~\ref{tab:const}.  Using the same values with the previous work makes easier to see the effect of the parameter transitions at radiation-matter equality, which was not taken into account in the previous work.

In the second case, we calculate the time evolution of $\gamma, \, \zeta, \, \bar{\Delta},$ and $\eta$ by solving the VOS equations \eqref{eq:veq} and \eqref{eq:gammaeq}.
The parameter values $\zeta, \eta, \bar{\Delta}$ are obtained from $v$ as
\begin{eqnarray}
  \zeta & = & \alpha \nu = (1-2v^2) \left (\frac{{\rm ln} \, (a/a_{\rm ini})}{{\rm ln} \, (t/t_{\rm ini})} \right ) \, , \label{eq:zetaeq}\\
    \eta & = & \frac{1}{2} cpv \, , \label{eq:etaeq}\\
    \bar{\Delta} & = & \frac{2 \pi}{35} \Biggl\{1+\frac{2}{3}(1-2v^2)-\frac{1}{11}(1-2v^2)^2 \Biggr\} \, . \label{eq:Deltaeq}
\end{eqnarray}
Table.~\ref{tab:const_scaling_solve} shows the asymptotic values of the parameters for RD and MD eras obtained by solving the VOS equations.  As we can find by comparing the two tables, some of the parameter values are different from the previous work, and they affect the kink distribution as well as the amplitude of the gravitational wave background.

\subsection{Gravitational waves from kinks}\label{sec:GWfromkink}
It has been shown in ref. \cite{Kawasaki:2010yi} that the kinks which contribute the most to the power of gravitational waves with angular frequency $\omega$ satisfy the following condition:
\begin{equation}
  \left (\psi \frac{N(\psi, \, t)}{V(t)/(\gamma t)^2} \right )^{-1} \sim \omega^{-1} . \label{eq:kink_N_omega_relation}
\end{equation}  
We define the sharpness of kinks which satisfies \eqref{eq:kink_N_omega_relation} for a given frequency $\omega$ as $\psi_{\rm max}(\omega, \, t)$.  This condition means that the main contribution on the gravitational wave background at physical frequency $\omega$ comes from kinks with sharpness $\psi_{\rm max}$ whose average\\ interval $(\psi N(\psi, \, t)/(V(t)/(\gamma t)^2))^{-1}$ is comparable with the wavelength of the gravitational waves $\omega^{-1}$. 

The strength of a gravitational wave burst from one kink on loops has been formalized in ref. \cite{Damour:2001bk}.  Including the dependence on the sharpness $\psi$, the strain amplitude is given by
\begin{equation} 
  h(f,z)=\frac{G\mu [\psi_{\rm max}(\omega,z)]^{1/2}l}{[(1+z)fl]^{2/3}}\frac{1}{r(z)} \Theta(1-\theta_m), \label{h}
\end{equation}
where $\theta_m = [(1+z)fl]^{-1/3}$, $f = a \omega/(2 \pi a_0)$ is the gravitational wave frequency today with $a_0=1$ being the present scale factor, $r$ is the distance to the source $r(z) = \int^z_{0} dz/H(z)$, and $l$ is twice the fundamental period $T_l=l/2$ of string loops. Since we consider infinite strings and their typical curvature is given by $\gamma t$, $l$ is replaced by the correlation length $\gamma t$ in our calculation. The step function $\Theta (1 -\theta_m)$ is introduced to set a low-frequency cutoff, which reflects the fact that kinks do not emit gravitational waves larger than the horizon size. We calculate the Hubble parameter using $H=H_0[\Omega_r (a/a_0)^{-4}+\Omega_m (a/a_0)^{-3}+\Omega_\Lambda]^{1/2}$, where $\Omega_r$, $\Omega_m$ and $\Omega_\Lambda$ are the density parameters for radiation, matter, and the cosmological constant, respectively.  We use $\Omega_r h^2=4.31\times 10^{-5}$ where $h$ is the reduced Hubble constant.  In this paper, we assume a flat universe and use the values obtained from Planck satellite \cite{Ade:2015xua}: $h=0.692$, $\Omega_m=0.308$ and $\Omega_\Lambda=0.692$. 

The power of the gravitational wave background is usually characterized by $\Omega_{\rm gw}\equiv ({\rm d}\rho_{\rm gw}/{\rm d}{\rm ln}f)/\rho_c $, where $\rho_{\rm gw}$ is the energy density of gravitational waves and $\rho_c$ is the critical density of the universe.  The gravitational wave spectrum generated from kinks on infinite strings is given by
\begin{equation}
  \Omega_{\rm gw}(f) = \frac{2\pi^2f^2}{3H_0^2}
  \int \frac{dz}{z}\Theta(n(f,z)-1) n(f,z)h^2(f,z) \, ,
  \label{eq:Omega_gw}
\end{equation}
where
\begin{equation}
  n(f,z)=\frac{1}{f}\frac{d\dot{N}}{d\ln z} \label{n} = \frac{1}{f}\cdot\frac{1}{2}\theta_m(f,z)\frac{z}{1+z}
  \frac{\psi_{\rm max}(\omega,z)N(\psi_{\rm max}(\omega,z),z)}{V}
  l^{-1}\frac{dV}{dz} \, ,
\end{equation}
and $dV/dz=4\pi a^3r^2(z)/H(z)$ is the volume between the redshift $z$ and $z + dz$.  The step function $\Theta(n(f,z)-1)$ is introduced to exclude rare bursts, whose intervals are longer than $\sim 1/f$ and cannot form a continuous background of gravitational waves.  Note the difference in the notation: $\psi\tilde{N}$ (number of kinks with sharpness $\ln\psi \sim \ln\psi + d\ln\psi$ per volume) in ref. \cite{Kawasaki:2010yi} is identical to $\psi N/V$ in our paper.  In summary, the differences with respect to ref. \cite{Kawasaki:2010yi} are 
\begin{itemize}
 \item We replace the typical curvature of infinite string as $l \sim \gamma t$ instead of $l\sim t$. 
 \item The probability of observing the gravitational wave burst from a kink is $\theta_m/2$ \cite{Olmez:2010bi} instead of $\theta_m/4$. 
 \item The distance $r$ and the volume $dV/dz$ are calculated numerically instead of using approximated analytic expressions. 
 \item $\Omega_\Lambda$ is included in the calculation of the Hubble parameter. 
\end{itemize}
These changes increase the overall spectral amplitude by $9.6$ in RD era and $2.7$ in MD era compared to the one calculated in ref. \cite{Kawasaki:2010yi}.

\section{Results}\label{sec:III}
\subsection{Result with the tanh interpolation}\label{sec:III_1}
\begin{figure}[htbp]
\centering
\includegraphics[width=12cm,clip]{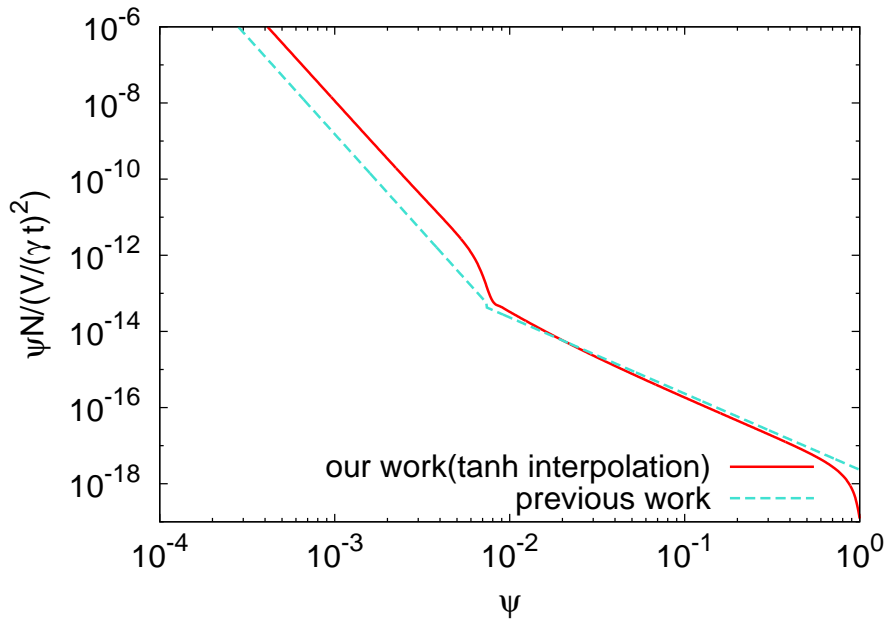}
\caption{The distribution function of kinks obtained using the tanh interpolation. 
The vertical axis is the number of kinks on infinite strings per length. The horizontal axis is the sharpness of kinks. The light-blue broken line is the analytic estimation in the previous work \cite{Kawasaki:2010yi} and the red solid line is our numerical result.}
\label{fig:Npsit0}
\end{figure}

We first solve the differential equation \eqref{eq:dNdtwithoutY} using the tanh interpolation \eqref{eq:tanh} with the values in table \ref{tab:const}.
The result is shown in figure~\ref{fig:Npsit0}. 
As mentioned in the previous section, the sharpness of kinks decreases with time. The number of old kinks with small sharpness is larger than new ones, because ${\cal O}(1-10)$ of kinks are produced per horizon and the number of newly produced kinks per comoving length decreases as the horizon grows. In figure~\ref{fig:Npsit0}, we find that the distribution function of kinks has two regions with different slopes. The left part ($\psi \lesssim 10^{-2}$) corresponds to kinks generated during the RD era, and the right part ($\psi \gtrsim 10^{-2}$) corresponds to kinks generated in the MD era.  Note that our result has a step at radiation-matter equality ($\psi \sim 7.4 \times 10^{-3}$), which is not seen in the result of the previous work. The reason will be discussed in the next section. 

\begin{figure}[htbp]
\centering
\includegraphics[width=12cm,clip]{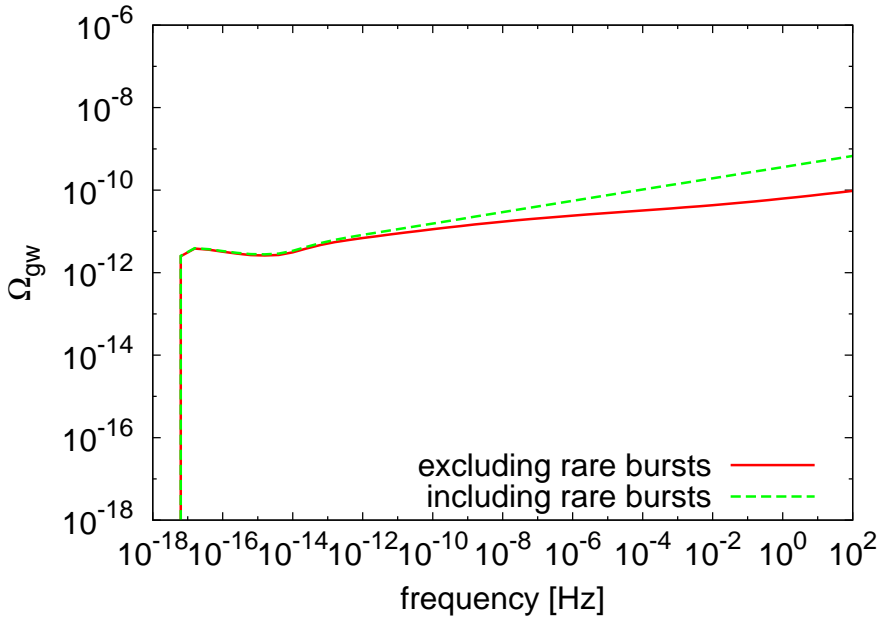}
\caption{Power spectrum of the gravitational wave background for $G \mu = 10^{-7}$. The red solid line is the result obtained excluding rare bursts and the green broken line is the result including rare bursts.}
\label{fig:Omega_gw} 
\end{figure}

Figure \ref{fig:Omega_gw} is the numerical results for the power spectrum of the gravitational wave background $\Omega_{\rm gw}$.
To calculate the gravitational wave background, we first look for the value which satisfies \eqref{eq:kink_N_omega_relation} each time in the calculation of $N(\psi,t)$ for each gravitational wave frequency $\omega$, and define it as $\psi_{\rm max}(\omega , t)$. Then, using the values of $\psi_{\rm max}$, we numerically integrate \eqref{eq:Omega_gw} to obtain the power spectrum.  Note that the vertical axis of figure~\ref{fig:Npsit0} is identical to the inverse of the left hand side of \eqref{eq:kink_N_omega_relation}.  So gravitational wave frequency $\omega$ is corresponded to the vertical axis of figure~\ref{fig:Npsit0}. 

As seen in figure~\ref{fig:Npsit0}, old kinks are numerous and the gravitational wave emission has a short interval, while new kinks are few and the interval is large. Thus, the high frequency gravitational waves are emitted from old kinks and low frequency gravitational waves are from new kinks. The gravitational waves in the range of $10^{-13} \, {\rm Hz} \leq f$ are generated from kinks with small sharpness produced in the RD era.  The middle frequency $10^{-15}  \, {\rm Hz} \leq f \leq 10^{-13} \, {\rm Hz}$ corresponds to kinks produced during the transition from the RD era to the MD era. The low frequency gravitational waves $f \leq 10^{-15} \, {\rm Hz}$ are emitted from kinks produced in the MD era.

\subsection{Result with the VOS model} \label{sec:III_2}
\begin{figure}[htbp]
    \centering
    \begin{tabular}{c}
      \begin{minipage}{0.5\hsize}
        \centering
        \includegraphics[width=7cm,clip]{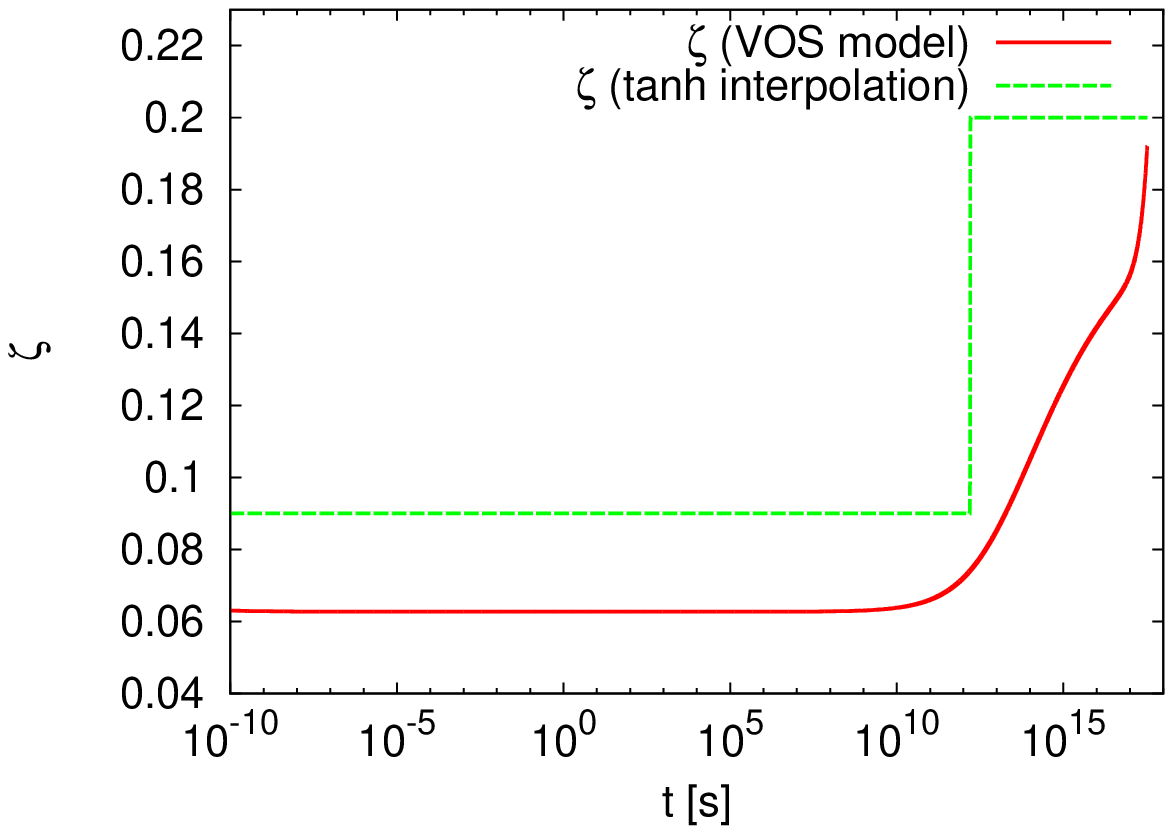}
      \end{minipage}
      \begin{minipage}{0.5\hsize}
        \centering
        \includegraphics[width=7cm,clip]{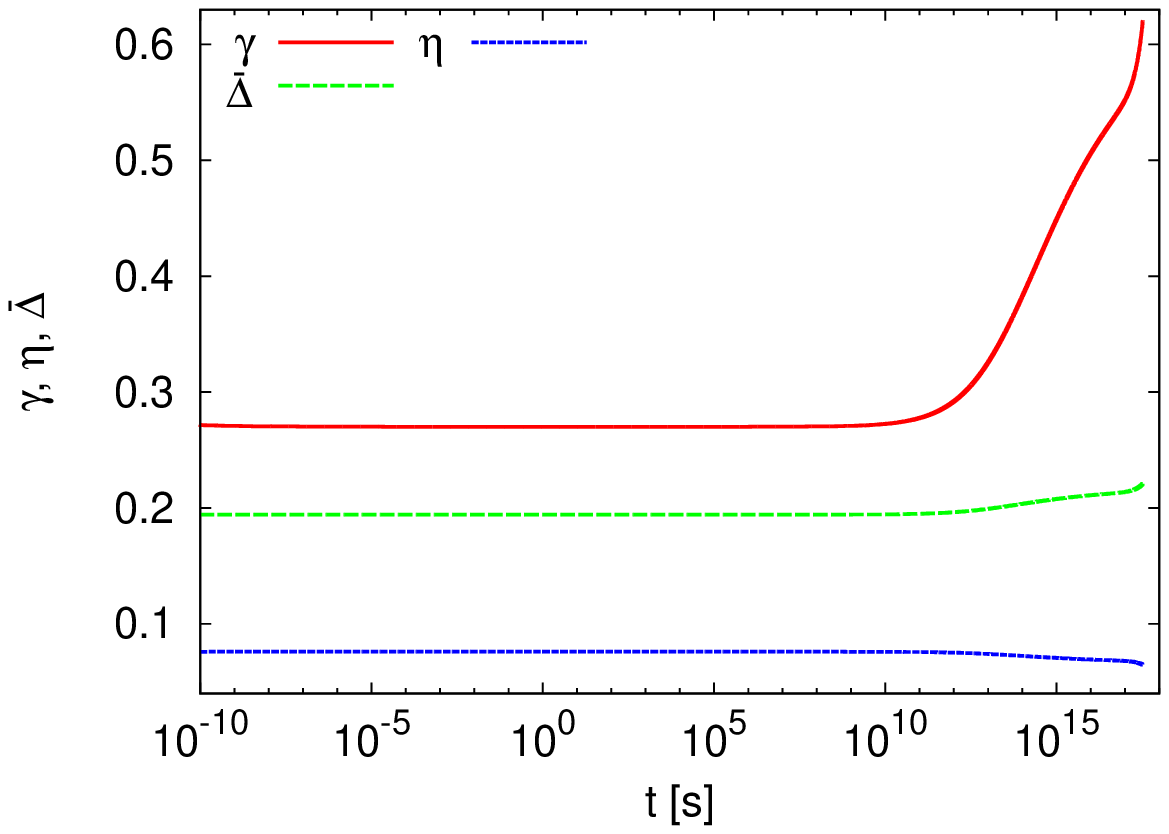}
      \end{minipage}
    \end{tabular}
    \caption{The time evolution of the parameters which characterize the production of kinks and their evolution. In the left figure, the time evolution of $\zeta$ is shown. The red solid line is obtained by solving VOS equations and the green broken line is the one interpolated by the tanh function. In the right figure, we show time evolution of $\gamma, \bar{\Delta}$ and $\eta$ obtained by solving the VOS equations. The red solid line is $\gamma$, the green broken line is $\bar{\Delta}$ and the blue broken line is $\eta$.}
    \label{fig:const}
  \end{figure}

In this section, we solve the differential equation \eqref{eq:dNdtwithoutY} by simultaneously solving the VOS equations \eqref{eq:veq} and \eqref{eq:gammaeq}. The VOS model provides time evolution of $\gamma$ and $v$, which can be converted to $\zeta, \, \eta$ and $\bar{\Delta}$ by \eqref{eq:zetaeq}, \eqref{eq:etaeq} and \eqref{eq:Deltaeq}. The time evolution of the parameters is shown in figure \ref{fig:const}.
We find their evolution is very different from the tanh interpolation. First, the VOS equations with $c=0.23$ provide different asymptotic values of the parameters as seen by comparing tables \ref{tab:const} and \ref{tab:const_scaling_solve}.  Second, the transition from the RD era to the MD era is not instant and it takes time to approach the asymptotic value.  In addition, the parameter values change near the present time, since we include cosmological constant.

\begin{figure}[htbp]
\centering
\includegraphics[width=12cm,clip]{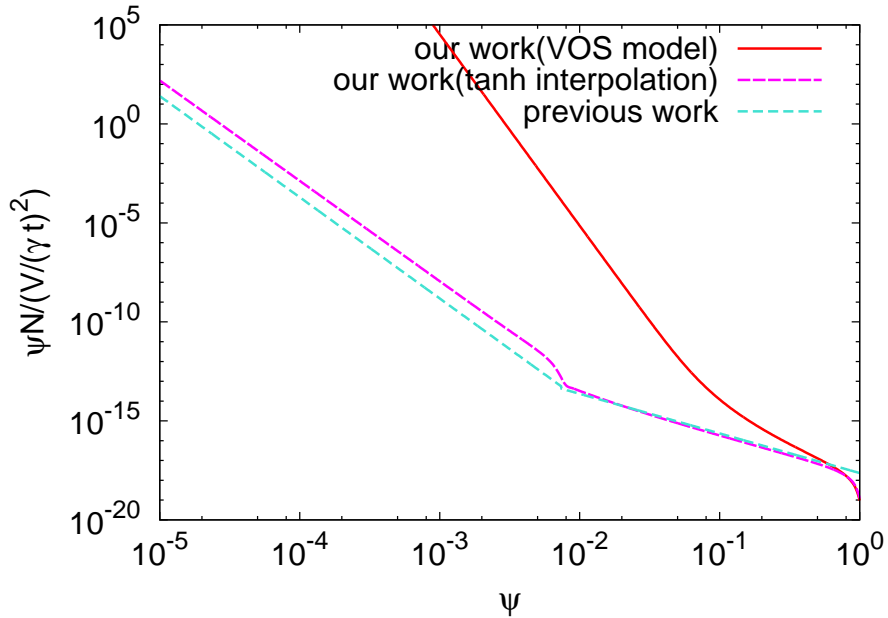}
\caption{The distribution function of kinks obtained by solving the VOS equations (the red solid line). The axises is same as figure \ref{fig:Npsit0}. For comparison, we also show the analytic estimation by the previous work \cite{Kawasaki:2010yi} (light-blue broken line) and our result of the tanh interpolation  (magenta broken line).}
\label{fig:Npsit_scaling} 
\end{figure}

The distribution of kinks obtained by the VOS model is shown in figure \ref{fig:Npsit_scaling}.
From the figure, we find two differences between the results with the tanh interpolation and the VOS model. First, the number of kinks increases considerably because the slope of the distribution function becomes steeper both for the RD and MD eras. Second, the position corresponding to radiation-matter equality has moved toward large $\psi$. The reason is discussed in the next section.

\begin{figure}[htbp]
\centering
\includegraphics[width=12cm,clip]{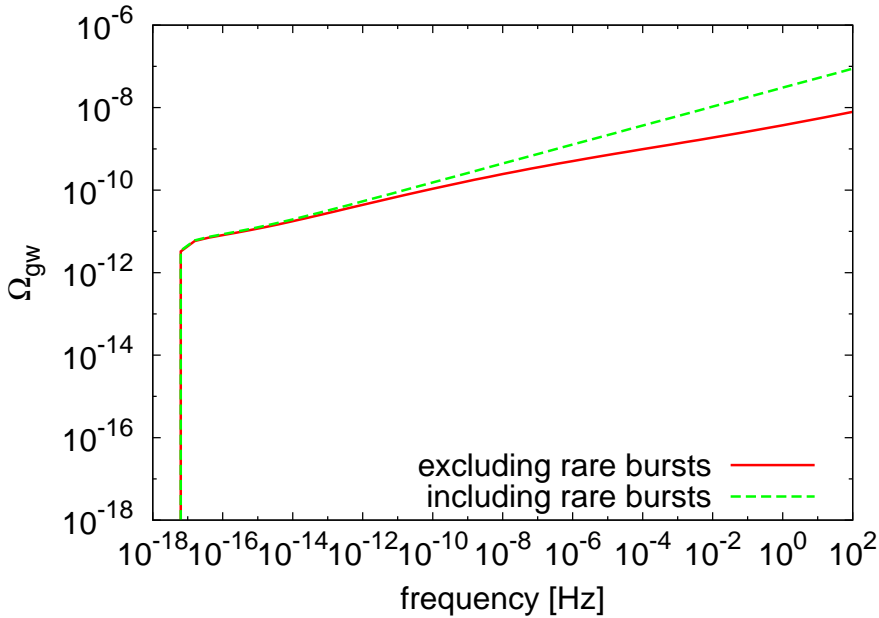}
\caption{Power spectrum of the gravitational wave background for $G \mu = 10^{-7}$. The axises are same as figure \ref{fig:Omega_gw}.}
\label{fig:Omega_gw_scaling} 
\end{figure}

Figure \ref{fig:Omega_gw_scaling} shows the power spectrum of the gravitational wave background $\Omega_{\rm gw}$ calculated using the kink distribution obtained by the VOS model. 
We see that the amplitude is larger than the case of the tanh interpolation, because of the increase in the number of kinks.  Since the number increases more at small $\psi$, which corresponds kinks generated during the RD era, the power of the gravitational wave spectrum is enhanced in high frequencies.

\begin{figure}[htbp]
\centering
\includegraphics[width=15cm,clip]{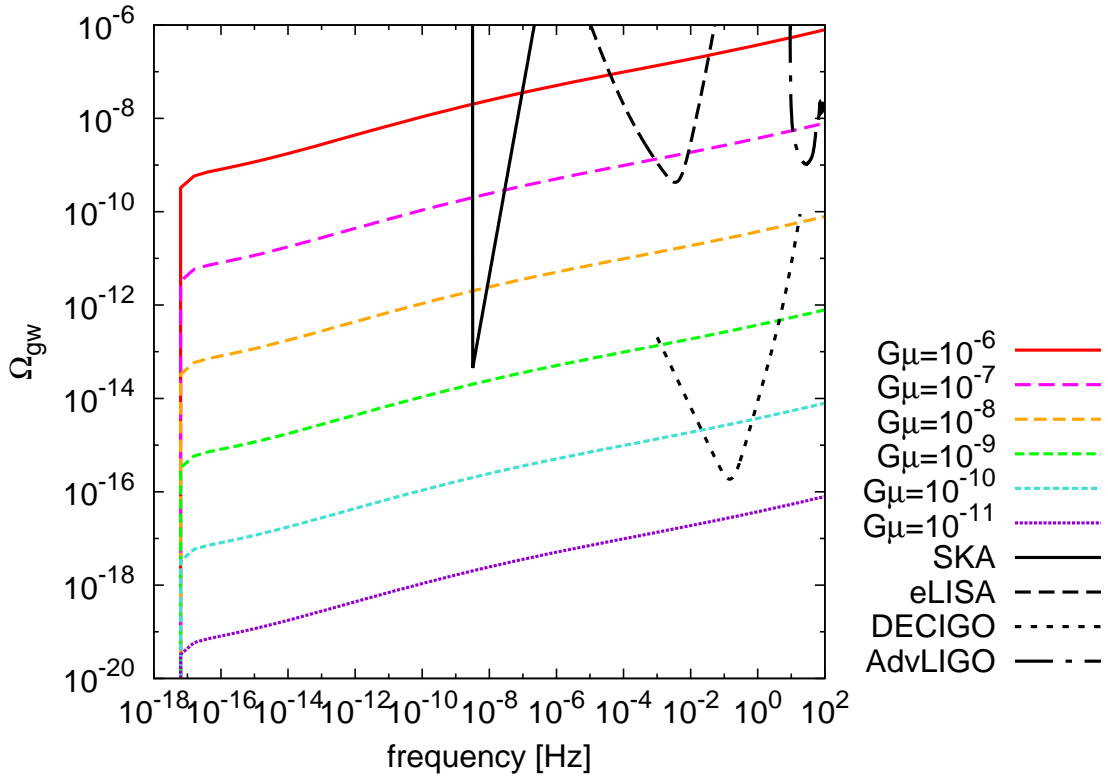}
\caption{The power spectrum of the gravitational wave background for different string tensions. The spectrum shown here do not include rare bursts. The black solid and broken lines are sensitivity curves of gravitational wave experiments.}
\label{fig:Omega_gw_scaling_iroiro} 
\end{figure}

Figure \ref{fig:Omega_gw_scaling_iroiro} is the comparison between sensitivity curves of future gravitational wave observations and the power spectra of the gravitational wave background for different values of string tension.  The SKA \cite{Janssen:2014dka} is a radio interferometer, which can detect gravitational waves by pulsar timing arrays. The eLISA \cite{AmaroSeoane:2012km,AmaroSeoane:2012je} and DECIGO \cite{Seto:2001qf,Kawamura:2011zz} missions will observe gravitational waves using laser interferometers at space.  Advanced-LIGO \cite{Harry:2010zz} is a laser interferometer constructed on the ground and will construct observation network with other ground-based detectors such as Advanced-VIRGO \cite{Accadia:2011zzc} and KAGRA in near future \cite{Somiya:2011np}.

\section{Discussion}\label{sec:IV}
First, let us compare our numerical result of the tanh interpolation with the previous work \cite{Kawasaki:2010yi}.  The major difference is that our result has a step-like feature in the distribution function of kinks at radiation-matter equality as seen in figure~\ref{fig:Npsit0}. This step arises because of the changes in the value of $\gamma$ in the evolution equation of the distribution function. 
The source term in  \eqref{eq:dNdtwithoutY} (the first term in the right hand side) has a factor of $\bar\Delta/\gamma^4$, and it becomes smaller in the MD era.  Thus, the number of newly produced kinks is smaller in the MD era.

The gravitational wave spectrum reflects the existence of this transition phase, and has three regions of different spectral slopes: the MD era $f \leq 10^{-15} \, {\rm Hz}$, the transition phase $10^{-15} \, {\rm Hz} \leq f \leq 10^{-13} \, {\rm Hz}$, and the RD era $10^{-13} \, {\rm Hz}\leq f$. Let us analytically estimate the frequency dependence of $\Omega_{\rm gw}$. 
The distribution function of kinks is related to the frequency of the gravitational wave background by \eqref{eq:kink_N_omega_relation}. The low-frequency gravitational waves $f \leq 10^{-15} \, {\rm Hz}$ are generated by kinks produced in the MD era, and we can read $\psi N/(V/(\gamma t)^2) \sim \psi^{-2}$ from figure~\ref{fig:Npsit0}. Then we can derive $\Omega_{\rm gw} \propto f^{-1/6}$ using \eqref{eq:kink_N_omega_relation} and \eqref{h}. This frequency dependence of the spectrum coincides with the analytic result in the previous work \cite{Kawasaki:2010yi} and is also consistent with the numerical result shown in figure~\ref{fig:Omega_gw}. The high-frequency region $10^{-13}  \, {\rm Hz}\leq f$ corresponds to the gravitational wave from kinks produced in the RD era, where the result in figure~\ref{fig:Npsit0} gives $ \psi N/(V/(\gamma t)^2)  \sim \psi^{-5.1}$ and we get $\Omega_{\rm gw} \propto f^{7/51}$. This also coincides with the analytic result of ref. \cite{Kawasaki:2010yi}, and is consistent with the spectrum with rare bursts in figure~\ref{fig:Omega_gw}. The difference from the previous work arises in $10^{-15}  \, {\rm Hz} \leq f \leq 10^{-13} \, {\rm Hz}$. In the transition phase, the value of $\psi_{\rm max}$ is the same for all the given frequency, so $\psi_{\rm max}\simeq {\rm Const.}$, and we get $\Omega_{\rm gw} \propto f^{1/3}$ from \eqref{eq:kink_N_omega_relation} and \eqref{h}. In fact, this frequency dependence can be seen in the numerical result of figure~\ref{fig:Omega_gw}.

Let us compare the spectral amplitude with the previous work \cite{Kawasaki:2010yi}.  We compare the results using the case with rare bursts, since the condition of excluding rare bursts depends on the distribution function of kinks and the comparison cannot be made simply.  First of all, the overall amplitude is $9.6$ and $2.7$ times larger than the previous work in RD and MD eras respectively, because of the modifications listed in section \ref{sec:GWfromkink}.  In addition, the amplitude of the high frequency region increases because of the larger number of kinks produced during the RD era. By extracting the dependence on $\psi_{\rm max}$ and $\gamma$ from \eqref{eq:Omega_gw}, we obtain
  \begin{eqnarray}
    \Omega_{\rm gw}(f) & \propto & f^2 \int \frac{{\rm d}z}{z} \psi_{\rm max}(\omega,z) \left(\psi_{\rm max}(\omega,z) \frac{N(\psi_{\rm max}(\omega,z),z)}{V/(\gamma t)^2} \right ) \gamma^{-8/3} \nonumber \\
    & \propto & f^2 \int \frac{{\rm d}z}{z} \psi_{\rm max}(\omega,z) \gamma^{-8/3} .
    \label{eq:Omega_gw_kantan}
  \end{eqnarray}
As seen in figure \ref{fig:Npsit0}, the number of kinks produced during RD era is larger than the previous work and the value of $\psi_{\rm max}$ becomes larger about twice for a fixed $\omega$. Therefore, for every frequency of the gravitational wave background corresponding to the RD era, the amplitude $\propto \psi_{\rm max}$ becomes twice larger than in the previous work. In total, the spectral amplitude is a few times larger in the low frequency and ${\mathcal O}(10)$ larger in the high frequency.

\vspace{5pt}
Next, let us discuss the case where we solve the parameter evolution with the VOS model. 
First, we explain why the distribution of kinks has different shape compared to the case of the tanh interpolation. In ref. \cite{Kawasaki:2010yi}, the solution of the distribution function is provided as 
  \begin{equation}
    \psi \frac{N}{V/(\gamma t)^2} \propto \psi^{\frac{-3+3\nu+\eta/\gamma}{2 \zeta}} .
    \label{eq:kink_bunpu_analysis}
  \end{equation}
As seen in tables \ref{tab:const} and \ref{tab:const_scaling_solve}, the parameter values from the VOS equations are different from the ones used in the tanh case. They largely affect the $\psi$ dependence of the distribution function as seen in \eqref{eq:kink_bunpu_analysis}. 
In particular, the small value of $\zeta$ increases the power of $\psi$, and makes the slope steeper. 
This increases the kink number considerably at small $\psi$.  In addition, the number of kink production is determined by the coeficient of $\bar{\Delta}/\gamma^4$ in the first term of the right hand side of \eqref{eq:dNdtwithoutY}.  The difference in this factor also increases the overall amplitude slightly.
In figure \ref{fig:Npsit_scaling}, we also see the position of radiation-matter equality shifts to larger $\psi$.
As provided in ref. \cite{Kawasaki:2010yi}, the value of $\psi$ corresponding to radiation-matter equality $\psi_*$ is given by 
  \begin{equation}
    \psi_*(t_0) = \left (\frac{t_{\rm eq}}{t_0} \right )^{2\zeta_m}, 
    \label{eq:rm-eq_position}
  \end{equation}
 where the suffix $m$ is the value during MD era. Since the value of $\zeta$ is smaller than the one used in the tanh case, $\psi_*$ becomes larger in the VOS case.

Then, let us describe the reason of the large increase of the spectral amplitude in figure \ref{fig:Omega_gw_scaling}.  The reason is the same as described in the tanh case, that is the increase of $\psi_{\rm max}$.  For example, $\psi_{\rm max}$ increases 100 times at $10^2$[Hz]. 
  Then, from \eqref{eq:Omega_gw_kantan}, $\Omega_{\rm gw}$ increases ${\mathcal O}(10^2)$.
  Taking account the enhancement of the overall amplitude $9.6$ for RD era, we find that the power spectrum has increased ${\mathcal O}(10^3)$ at high frequencies compared to the previous work.

From figure~\ref{fig:Omega_gw_scaling_iroiro}, we find that the gravitational wave background from kinks on infinite strings is testable by future experiments depending on the tension of strings. The SKA would probe $G \mu \gtrsim 10^{-9}$, eLISA and Advanced-LIGO can test $G \mu \gtrsim 10^{-7}$, and DECIGO has the strongest sensitivity to reach $G \mu \gtrsim 10^{-10}$.  Note that the spectrum shown in figure~\ref{fig:Omega_gw_scaling_iroiro} do not include rare bursts.  Since rare bursts with large amplitude exists at high frequencies as seen in figure~\ref{fig:Omega_gw}, laser-interferometer experiments could be also used to search for rare burst signals.

\section{Conclusions}\label{sec:V}
In this work, we have calculated the power spectrum of the gravitational wave background from kinks on infinite strings.  First, we have solved the differential equation \eqref{eq:dNdtwithoutY} to obtain the distribution function of kinks numerically in two ways. First, unlike the analytic estimation of
ref. \cite{Kawasaki:2010yi}, we have smoothly connected the parameters using a tangent hyperboric function, which are related to the evolution of the cosmic string network, at radiation-matter equality. As a result, we have found a step in the distribution function of kinks at the transition from the RD era to the MD era, which was overlooked in the previous work \cite{Kawasaki:2010yi}.
At the same time, we have found an increase in the number of kinks generated in the RD era. Second, we have calculated the distribution of kinks by following the time evolution of the parameters with the VOS equations. We have found a steeper slope of the distribution function, which gives a large increase of the kink number at small sharpness, and the shift of the position of radiation-matter equality.

Next, using the numerical result of the distribution function of kinks, we have calculated the power spectrum of the gravitational wave background.
In the case where we use the tanh interpolation, due to the step in the distribution function of kinks, we have found that the power spectrum behaves as $\Omega_{\rm gw} \propto f^{1/3}$ at $10^{-15}  \, {\rm Hz} \lesssim f \lesssim 10^{-13} \, {\rm Hz}$. The power spectrum has increased more in the case where we solve the the VOS equations. In addition to the precise estimation of the kink distribution, we have also carefully evaluated all the factors involved in the calculation of the spectrum.  This allows us to offer a rather precise prediction on the spectral amplitude.
By comparing the results with sensitivities of future experiments, 
we have shown that gravitational waves from kinks on infinite strings can be probed at different frequencies.

Finally, let us comment on the gravitational wave background from cusps and kinks on cosmic string loops. Loops emit gravitational waves whose wavelength is shorter than the loop size, and usually the number of loops is more than that of infinite strings. Thus, it is more likely that the gravitational wave background generated by infinite strings is sub-dominant at high frequencies compared to the one from loops.  However, since the typical loop size is not known yet, gravitational waves from kinks on infinite strings could be more important than that from loops, especially for the SKA which probes low frequencies. It is important to consider both origins of gravitational wave background to provide constraints on cosmic strings, and thus our careful estimation of the gravitational wave background from kinks on infinite strings would help to constrain cosmic strings by observation at low frequencies.
Moreover, there are cosmic superstrings predicted in superstring theory, and they also form the network consisting of loops and infinite strings. They have cusps and kinks which emit gravitational waves.
  Recently, the power spectrum of the gravitational wave background produced by loops of cosmic superstrings has been investigated \cite{Sousa:2016ggw}, but the one from kinks on infinite superstrings is not clear yet.  
It is being examined in a work in progress.
If a new era of multi-wavelength gravitational wave observations is successful and a detection was made, we might even be able to get insight in the physics of the very early universe.

\acknowledgments
KH is supported by a Grant-in-Aid for JSPS Research under Grant No.15J05029,
DN is supported by MEXT Grant-in-Aid for Scientific Research on Innovative Areas,~No.15H05890, 
SK is supported by Career Development Project for Researchers of Allied Universities.

\end{document}